\documentclass[12pt]{article} 
\usepackage{epsfig,amsfonts,amssymb}
 
\newcommand{\be}{\begin{equation}} 
\newcommand{\ee}{\end{equation}}
\newcommand{\bea}{\begin{eqnarray}} 
\newcommand{\eea}{\end{eqnarray}}
\newcommand{\beas}{\begin{eqnarray*}} 
\newcommand{\eeas}{\end{eqnarray*}}
\newcommand{\btheta}{\vartheta}

\begin{document}

\voffset -0.7 true cm \hoffset 1.1 true cm \topmargin 0.0in
\evensidemargin 0.0in \oddsidemargin 0.0in \textheight 8.7in
\textwidth 7.3in \parskip 10 pt

\begin{titlepage}
  \begin{flushright}
    {\small TIFR/TH/05-47} \\
    {\small hep-th/0512138}
  \end{flushright}

  \begin{center}
    \vspace{26mm}

    {\LARGE \bf A C-Function For Non-Supersymmetric Attractors}

    \vspace{10mm}

    Kevin Goldstein,  Rudra P. Jena, Gautam Mandal and Sandip P.
    Trivedi

    \vspace{5mm}

    {\small \sl Tata Institute of Fundamental Research} \\
    {\small \sl Homi Bhabha Road, Mumbai, 400 005, INDIA} \\
    {\small \tt kevin,  rpjena, mandal, sandip@theory.tifr.res.in}
    \vspace{10mm}

  \end{center}

  \vskip 0.3 cm \centerline{\bf Abstract} \vspace{5mm}
  \noindent
  We present a c-function for spherically symmetric, static and
  asymptotically flat solutions in theories of four-dimensional
  gravity coupled to gauge fields and moduli.  The c-function is valid
  for both extremal and non-extremal black holes.  It monotonically
  decreases from infinity and in the static region acquires its
  minimum value at the horizon, where it equals the entropy of the
  black hole.  Higher dimensional cases, involving $p$-form gauge
  fields, and other generalisations are also discussed.
 
\end{titlepage}
\section{Introduction}

The attractor phenomenon for extremal black holes has been the subject of considerable
investigation.  For BPS black holes in ${\cal N}=2$ theories this phenomenon was first
studied in \cite{Ferrara:1995ih} and thereafter discussed in
\cite{Cvetic:1995bj,Strominger:1996kf,Ferrara:1996dd,Ferrara:1996um,Cvetic:1996zq,
  Ferrara:1997tw,Gibbons:1996af,Denefa,Denef:2001xn}.
It has received further attention recently due to the conjecture of \cite{Ooguri:2004zv}
and related developments
\cite{LopesCardoso:1998wt, LopesCardoso:1999ur, Dabholkar:2004yr,Ooguri:2005vr,Dijkgraaf:2005bp}.  For
non-supersymmetric extremal black holes, some aspects of the attractor phenomenon were
discussed in \cite{Ferrara:1997tw} and \cite{Gibbons:1996af}. More recently this has been
investigated in \cite{Sen:2005wa} and \cite{gijt}, \cite{Kallosh:2005bj},
\cite{Kallosh:2005ax}, \cite{TTNSA}, \cite{Giryavets:2005nf}.  For important related work
see \cite{Kraus:2005vz}, \cite{Kraus:2005zm}.

In supersymmetric black holes, the central charge, which is a function of the moduli and
the charges carried by the black hole, plays an important role in the discussion of the
attractor.  The attractor values of the scalars, which are obtained at the horizon of the
black hole, are given by minimising the central charge with respect to the moduli.  In the
non-supersymmetric case one constructs an effective potential which is a function of the
moduli and charges. The attractor values are then given by minimising this effective
potential with respect to the moduli.

There is a another sense in which the central charge is also minimised at the horizon of a
supersymmetric attractor.  One finds that the central charge, now regarded as a function
of the position coordinate, evolves monotonically from asymptotic infinity to the horizon
and obtains its minimum value at the horizon of the black hole.  It is natural to ask
whether there is an analogous quantity in the non-supersymmetric case and in particular if
the effective potential is also monotonic and minimised in this sense for
non-supersymmetric attractors.

This paper addresses this question.  We present a c-function for non-supersymmetric
attractors here. We first study the four dimensional case.  The c-function has a simple
geometrical and physical interpretation in this case. We are interested in spherically
symmetric and static configurations in which all fields are functions of only one variable
- the radial coordinate.  The c-function, $c(r)$, is given by
\be
\label{defcintro}
c(r)={1\over 4} A(r), 
\ee
where $A(r)$ is the area of the two-sphere, of the $SO(3)$ isometry group orbit, as a
function of the radial coordinate \footnote{We have set $G_N=1$.}.  For any asymptotically
flat solution we show that the area function satisfies a c-theorem and monotonically
decreases as one moves inwards from infinity.  For a black hole solution, the static
region ends at the horizon, so in the static region the c-function attains its minimum
value at the horizon. This horizon value of the c-function equals the entropy of the black
hole. While the horizon value of the c-function is also proportional to the minimum value
of the effective potential, more generally, away from the horizon, the two are different.
In fact we find that the effective potential need not vary monotonically in a
non-supersymmetric attractor.  The c-theorem we prove is applicable for supersymmetric
black holes as well.  In the supersymmetric case, there are three quantities of interest,
the c-function, the effective potential and the square of the central charge. At the
horizon these are all equal, up to a constant of proportionality.  But away from the
horizon they are in general different.

We work directly with the second order equations of motion in our analysis and it might
seem puzzling at first that one can prove a c-theorem at all. The answer to the puzzle
lies in boundary conditions. For black hole solutions we require that the solutions are
asymptotically flat. This is enough to ensure that going inwards from asymptotic infinity
the c-function decreases monotonically.  Without imposing any boundary conditions one
cannot prove the c-theorem, as one might expect. But one can show that in the absence of
singularities, $c$ can have at most one critical point.

While non-supersymmetric attractors were our primary motivation, the c-theorem is in fact
valid for all static, spherically symmetric, asymptotically flat, solutions to the
equations of motion
\footnote{Also there the spacetime region under consideration must be singularity free.}.
For example, the proof applies also to non-extremal black holes. Once again the Area
function must decreases monotonically and its minimum value at the horizon is the entropy.

In our discussion we focus on a system consisting of 4-dimensional gravity coupled to
gauge fields and moduli.  But in fact the results are more general. The c-theorem is valid
for any matter fields which satisfy the null energy condition. This says that, 
\be
\label{wecond}
T_{\mu\nu}\zeta^\mu\zeta^\nu \geqslant 0, 
\ee 
for any null vector $\zeta$.  As long as this energy condition is met and we have a
static, spherically symmetric solution that is asymptotically flat, the area function
monotonically decreases, moving in from infinity.  The importance of the null energy
condition in the proof of a c-theorem was recognised in \cite{fgpw}.
 
One can show that the proof of the c-theorem follows in a straightforward manner from the
Raychaudhuri equation and the energy condition, eq.(\ref{wecond}). By considering a
congruence of radially infalling null geodesics one can see that the area $A(r)$ must
decrease as one moves inwards from asymptotic infinity.  Our focus here is on spherically
symmetric configurations, but these comments suggest that a similar c-theorem can be
devised more generally as well.
 
In the latter part of this paper we consider generalisations to higher dimensions.  We
analyse a system of rank $q$ gauge fields and moduli coupled to gravity and once again
find a c-function that satisfies a c-theorem. In $D=p+q+1$ dimensions this system has
extremal black brane solutions whose near horizon geometry is $AdS_{p+1} \times S^q$.  We
show that the c-function is non-increasing from infinity up to the near horizon region.
It's minimum value in the $AdS_{p+1} \times S^q$ region agrees with the conformal anomaly
in the dual boundary theory for $p$ even. A c-function in $AdS$ space was considered
before in \cite{fgpw},\cite{Girardello:1998pd} and our construction makes important use of the analysis and results
contained therein.

In fact, in the higher dimensional case as well, the c-theorem we prove is more general.
It applies to all solutions which have a $SO(q) \times P$ symmetry, where $P$ is the
Poincare group in $p+1$ dimensions, as long as suitable boundary conditions are imposed.
Both asymptotically flat and asymptotically $AdS$ boundary conditions lead to
monotonicity.  And both extremal and non-extremal black brane solutions are examples which
satisfy the conditions for the $c$-theorem. Also, the c-theorem works for other kinds of
matter we well, as long as the null energy condition holds.

This paper is structured as follows. In Section 2, we discuss some background material.
Section 3, discusses the $c$-theorem in $4$ dimensions and Section 4, the higher
dimensional case. Three appendices contain important details.

\section{Background}

We begin with some background related to the discussion of non-supersymmetric attractors.

Consider a theory consisting of four dimensional gravity coupled to $U(1)$ gauge fields
and moduli, whose bosonic terms have the form,
\begin{equation}
  S=\frac{1}{\kappa^{2}}\int d^{4}x\sqrt{-G}(R-2 g_{ij}(\partial\phi^i) (\partial \phi^j)-
  f_{ab}(\phi^i)F^a_{\mu \nu} F^{b \ \mu \nu} -{\textstyle{1 \over 2}} {\tilde f}_{ab}(\phi^i) F^a_{\mu \nu}
  F^b_{\rho \sigma} \epsilon^{\mu \nu \rho \sigma} ).
  \label{actiongen}
\end{equation}
$F^a_{\mu\nu}, a=0, \cdots N$ are gauge fields.  $\phi^i, i=1, \cdots n$ are scalar
fields.  The scalars have no potential term but determine the gauge coupling constants.
We note that $g_{ij}$ refers to the metric in the moduli space, this is different from the
spacetime metric, $G_{\mu\nu}$.

A spherically symmetric space-time metric in $3+1$ dimensions takes the form,
\begin{eqnarray}
  ds^{2} & = & -a(r)^{2}dt^{2}+a(r)^{-2}dr^{2}+b(r)^{2}d\Omega^{2}\label{metric2}
\end{eqnarray}

The Bianchi identity and equation of motion for the gauge fields can be solved by a field
strength of the form,
\begin{equation}
  \label{fstrengthgen}
  F^a=f^{ab}(Q_{eb}-{\tilde f}_{bc}Q^c_m) {1\over b^2} dt\wedge dr + Q_m^a sin \theta  d\theta \wedge d\phi,
\end{equation}
where $Q_m^a, Q_{ea}$ are constants that determine the magnetic and electric charges
carried by the gauge field $F^a$, and $f^{ab}$ is the inverse of $f_{ab}$.

The effective potential $V_{eff}$ is then given by,
\begin{equation}
  \label{defpotgen}
  V_{eff}(\phi_i)=f^{ab}(Q_{ea}-{\tilde f}_{ac}Q^c_m)(Q_{eb}- {\tilde f}_{bd}Q^d_m)+f_{ab}Q^a_mQ^b_m.
\end{equation}

For the attractor mechanism it is sufficient that two conditions to be met.  First, for
fixed charges, as a function of the moduli, $V_{eff}$ must have a critical point.
Denoting the critical values for the scalars as $\phi^i=\phi^i_0$ we have,
\begin{equation}
  \label{critical}
  \partial_iV_{eff}(\phi^i_{0})=0.
\end{equation}
Second, the effective potential must be a minimum at this critical point.  I.e. the matrix
of second derivatives of the potential at the critical point,
\begin{equation}
  \label{massmatrix}
  M_{ij}={1\over 2} \partial_i\partial_jV_{eff}(\phi^i_{0})
\end{equation}
should have positive eigenvalues.  Schematically we can write,
\begin{equation}
  \label{positive}
  M_{ij}>0.
\end{equation}
As discussed in \cite{TTNSA}, it is possible that some eigenvalues of $M_{ij}$ vanish. In
this case the leading correction to the effective potential along the zero mode directions
should be such that the critical point is a minimum. Thus, an attractor would result if
the leading correction is a quartic term,
$V_{eff}=V_{eff}(\phi^i_0) + \lambda (\phi -\phi_H)^4$, with $\lambda >0$ but not if it is
a cubic term, $V_{eff}=V_{eff}(\phi^i_0) + \lambda (\phi -\phi_H)^3$.

Once the two conditions mentioned above are met it was argued in \cite{gijt} that the
attractor mechanism works.  There is an extremal Reissner Nordstrom black hole solution in
the theory, where the black hole carries the charges specified by the parameters,
$Q^a_m, Q_{ea}$ and the moduli take the critical values, $\phi_0$ at infinity.  For small
enough deviations at infinity of the moduli from these values, a double-horizon extremal
black hole solution continues to exist. In this extremal black hole the scalars take the
same fixed values, $\phi_0$, at the horizon independent of their values at infinity.  The
resulting horizon radius is given by,
\begin{equation}
  \label{RH}
  b_H^2=V_{eff}(\phi^i_{0})
\end{equation}
and the entropy is \be
\label{BH}
S_{BH}={1 \over 4} A = \pi b_H^2. \ee

In ${\cal N}=2$ supersymmetric theory, $V_{eff}$ can be expressed, \cite{Ferrara:1997tw},
in terms of a Kahler potential, $K$ and a superpotential, $W$ as,
\begin{equation}
  \label{epot}
  V_{eff}=e^K[g^{i \bar j}\nabla_i W (\nabla_j W)^* + |W|^2],
\end{equation}
where $\nabla_iW\equiv \partial_iW+\partial_iK W$.  The Kahler potential and
Superpotential in turn can be expressed in terms of a prepotential $F$, as,
\begin{equation}
  \label{kpsg}
  K=-\ln Im(\sum_{a=0}^N X^{a*}\partial_aF(X)),
\end{equation}
and,
\begin{equation}
  \label{superpotsg}
  W=q_aX^a - p^a \partial_aF,
\end{equation}
respectively. Here, $X^a, a=0, \cdots N$ are special coordinates to describe the special
geometry of the vector multiplet moduli space.  And $q_a,p^a$ are the electric and
magnetic charges carried by the black hole
\footnote{These can be related to $Q_{ea},Q^a_m,$ using eq.(\ref{fstrengthgen}).}.

For a BPS black hole, the central charge given by,
\begin{equation}
  \label{ccharge}
  Z=e^{K/2}W, 
\end{equation}
is minimised, i.e., $\nabla_iZ=\partial_iZ+{1\over 2} \partial_iK Z=0$. This condition is
equivalent to,
\begin{equation}
  \label{attractorsusy}
  \nabla_iW=0.
\end{equation}
The resulting entropy is given by
\begin{equation}
  \label{susyentropy}
  S_{BH}=\pi e^K |W|^2.
\end{equation}
with the Kahler potential and superpotential evaluated at the attractor values.

\section{The c-function in $4$ Dimensions. }
\subsection{The c-function}
The equations of motion which follow from eq.(\ref{actiongen}) take the form,
\begin{equation}
  \label{motion}
  \begin{array}{rcl}
    R_{\mu\nu}-2g_{ij} \partial_{\mu}\phi^i\partial_{\nu}\phi^j
    & = &   f_{ab} \left(2F^a_{\phantom{a}\mu\lambda}F^{b\phantom{\nu}\lambda}_{\phantom{b}\nu}-
      {\textstyle \frac{1}{2}}G_{\mu\nu}F^a_{\phantom{a}\kappa\lambda} F^{b \kappa\lambda} \right)  \\
    \frac{1}{\sqrt{-G}}\partial_{\mu}\left(\sqrt{-G}g_{ij}\partial^{\mu}\phi^j\right)
    & = & {\textstyle{1 \over 4}} \partial_i(f_{ab}) F^a _{\phantom{a}\mu\nu} F^{b \mu\nu} \\
    && + {\textstyle{1\over 8}} \partial_i({\tilde f}_{ab}) F^a_{\mu\nu} F^b_{\rho \sigma} \epsilon^{\mu\nu\rho\sigma}  \\
    \partial_{\mu}\left(\sqrt{-G}(f_{ab} F^{b \mu\nu} + {\textstyle{1\over 2}} {\tilde f}_{ab}F^b_{\rho\sigma}
      \epsilon^{\mu\nu\rho\sigma} )\right) & = & 0. 
  \end{array}
\end{equation}

We are interested in static, spherically symmetric solutions to the equations of motion.
The metric and gauge fields in such a solution take the form, eq.(\ref{metric2}),
eq.(\ref{fstrengthgen}).  We will be interested in asymptotically flat solutions below.
For these the radial coordinate $r$ in eq.(\ref{metric2}) can be chosen so that
$r\rightarrow \infty$ is the asymptotically flat region.

The scalar fields are a function of the radial coordinate alone, and substituting for the
gauge fields from, eq.(\ref{fstrengthgen}), the equation of motion for the scalar fields
take the form,
\begin{equation}
  \label{eomdil}
  \partial_{r}(a^{2}b^{2}g_{ij}\partial_{r}\phi^j)=\frac{\partial_iV_{eff} }{2b^{2}},
\end{equation}
where $V_{eff}$ is defined in eq.(\ref{defpotgen}).
 
The Einstein equation for the $rr$ component takes the form of an ``energy constraint'',
\begin{equation}
  -1+a^{2}b^{'2}+\frac{a^{2'}b^{2'}}{2}=\frac{-1}{b^{2}}(V_{eff}(\phi_i))+a^{2}b^{2}
  g_{ij} (\partial_{r}\phi^i) \partial_r\phi^j \label{constraint}
\end{equation}

Of particular relevance for the present discussion is the equation obtained for
$R_{rr}-{G^{tt}\over G^{rr}} R_{tt}$ component of the Einstein equation. From
eq.(\ref{motion}), this is,
\begin{equation}
  \frac{b(r)^{''}}{b(r)}=-g_{ij}\partial_{r}\phi^i \partial_r\phi^j. \label{eq2}
\end{equation}
Here prime denotes derivative with respect to the radial coordinate $r$.

Our claim is that the c-function is given by,
\begin{equation}
  \label{cfunc4}
  c={1\over 4} A(r),
\end{equation}
where $A(r)$ is the area of the two-sphere defined by constant $t$ and $r$,
\begin{equation}
  \label{area}
  A(r)=\pi b^2(r).
\end{equation}

We show below that in any static, spherically symmetric, asymptotically flat solution, $c$
decreases monotonically as we move inwards along the radial direction from infinity.  We
assume that the spacetime in the region of interest has no singularities and the scalar
fields lie in a singularity free region of moduli space with a metric which is positive,
i.e., all eigenvalues of the moduli space metric, $g_{ij}$, are positive.  For a black
hole we show that the minimum value of $c$, in the static region, equals the entropy at
the horizon.

To prove monotonicity of $c$ it is enough to prove monotonicity of $b$.  Let us define a
coordinate $y=-r$ which increases as we move inwards from the asymptotically flat region.
We see from eq.(\ref{eq2}), since the eigenvalues of $g_{ij}>0$, that
$d^2b/dy^2 \leqslant 0$ and so $db/dy$ must be non-increasing as $y$ increases.  Now for
an asymptotically flat solution, at infinity as $r \rightarrow \infty$,
$b(r)\rightarrow r$.  This means $db/dy=-1$.  Since $db/dy$ is non-increasing as $y$
increases this means that for all $y>-\infty$, $db/dy<0$ and thus $b$ is monotonic.  This
proves the c-theorem.

\subsection{Some Comments}

A few comments are worth making at this stage.

It is important to emphasise that our proof of the c-theorem applies to any spherically
symmetric, static solution which is asymptotically flat.  This includes both extremal and
non-extremal black holes. The boundary of the static region of spacetime, where the
killing vector ${\partial \over \partial t }$ is time-like, is the horizon where
$a^2\rightarrow 0$. The $c$ function is monotonically decreasing in the static region, and
obtains its minimum value on the boundary at the horizon.  We see that this minimum value
of $c$ is the entropy of the black hole.  We will comment on what happens to $c$ when one
goes inside the horizon towards the end of this section.

For extremal black holes it is worth noting that the c-function is not $V_{eff}$ itself.
At the horizon, where $c$ obtains its minimum value, the two are indeed equal (up to a
constant of proportionality).  This follows from the constraint, eq.(\ref{constraint}),
after noting that at a double horizon where $a^2 $ and $a^{2'}$ both vanish,
$V_{eff}(\phi^i_0)=b_H^2$. But more generally, away from the horizon, $c$ and $V_{eff}$
are different. In particular, we will consider an explicit example in appendix
\ref{sec:v_eff-need-not} of a flow from infinity to the horizon where $V_{eff}$ does not
evolve monotonically.

In the supersymmetric case it is worth commenting that the c-function discussed above and
the square of the central charge agree, up to a proportionality constant, at the horizon
of a black hole.  But in general, away from the horizon, they are different.  For example
in a BPS extremal Reissner Nordstrom black hole, obtained by setting the scalars equal to
their attractor values at infinity, the central charge is constant, while the Area is
infinite asymptotically and monotonically decreases to its minimum at the horizon.

It is also worth commenting that $c'$ can vanish identically only in a Robinson-Bertotti
spacetime
\footnote{By $c'$ vanishing identically we mean that $c'$ and all its derivatives vanish
  in some region of spacetime.}.
If $c$ is constant, $b$ is constant. From, eq.(\ref{eq2}) then $\phi^i$ are constant.
Thus $V_{eff}$ is extremised. It follows from the other Einstein equations then that
$a(r)=r/b$ leading to the Robinson-Bertotti spacetime.  From this we learn that a flow
from one asymptotically (in the sense that $c'$ and all its derivatives vanish)
$AdS_2 \times S^2$ where the scalars are at one critical point of $V_{eff}$ to an
asymptotically $AdS_2\times S^2$ spacetime where the scalars are at another critical point
is not possible.  Once the scalars begin evolving $c'$ will became negative and cannot
return to zero.
   
The c-theorem discussed above is valid more generally than the specific system consisting
of gravity, gauge fields and scalars we have considered here.  Consider any
four-dimensional theory with gravity coupled to matter which satisfies the null energy
condition.  By this we mean that the stress-energy satisfies the condition, 
\be
\label{wec}
T_{\mu\nu} \zeta^\mu\zeta^\nu \geqslant 0, 
\ee 
where $\zeta^a$ is an arbitrary null
vector.  One can show that in such a system the c-theorem is valid for
all static, spherically symmetric, asymptotically flat, solutions of
the equations of motion. To see this, note that from the metric
eq.(\ref{metric2}), it follows that, 
\be
\label{gene}
-R_{tt}G^{tt}+R_{rr}G^{rr}= -2 a^2 {b^{''} \over b}.  
\ee 
From Einstein's equations and the null energy condition we learn that the
l.h.s above is positive, since 
\be
\label{conda}
-R_{tt}G^{tt}+R_{rr}G^{rr}=T_{\mu\nu}\zeta^\mu\zeta^\nu >0 
\ee 
where $\zeta^\mu=(\zeta^t,\zeta^r)$ are components of a null vector, satisfying the
relations, $(\zeta^t)^2=-G^{tt}, (\zeta^r)^2=G^{rr}$.  Thus as long as we are outside the
horizon, and $a^2>0$, i.e. in any region of space-time where the Killing vector related to
time translations is time-like, $b^{''}<0$
\footnote{In fact the same conclusion also holds inside the horizon.  Now $t$ is
  space-like and $r$ time-like and
  $T_{\mu\nu}\zeta^\mu\zeta^\nu=2a^2 {b^{''}\over b}\geqslant 0$.  Since $a^2<0$, we conclude
  that $b^{''}<0$. We will return to this point at the end of the section.}.
This is enough to then prove the monotonicity of $b$ and thus $c$.  The importance of the
null energy condition for a c-theorem was emphasised in \cite{fgpw}
\footnote{In \cite{fgpw} this condition is referred to as the weaker energy condition.}.

In fact the c-theorem follows simply from the Raychaudhuri equation and the null energy
condition.  Consider a congruence of null geodesics, where each geodesic has
$(\theta, \phi)$ coordinates fixed, with, $(t, r)$, being functions of the affine
parameter, $\lambda$.  The expansion parameter of this congruence is
\be
\label{expan}
{\btheta}={d \ln A \over d\lambda}, 
\ee 
where $A$ is the area, eq.(\ref{area}).  Choosing in going null geodesics for which
$dr/d\lambda<0$ we see that $\btheta<0$ at $r \rightarrow \infty$, for an asymptotically
flat space-time.  Now, Raychaudhuri's equation tells us that ${d\btheta \over d\lambda}<0$
if the energy condition, eq.(\ref{wec}), is met.  Then it follows that $\btheta <0$ for
all $r<\infty$ and thus the area $A$ must monotonically decrease.  The comments in this
paragraph provides a more coordinate independent proof of the c-theorem. Although the
focus of this paper is time independent, spherically symmetric configurations, these
comments also suggest that a similar c-theorem might be valid more generally.  The
connection between c-theorems and the Raychaudhuri equation was emphasised in
\cite{Sahakian:1999bd}, \cite{Alvarez:2000jb}.

In the higher dimensional discussion which follows we will see that
the $c$ function is directly expressed in terms of the expansion
parameter $\btheta$ for radial null geodesics. The reader might wonder
why we have not considered an analogous $c$ function in
four-dimensions. From the discussion of the previous paragraph we see
that any function of the form, $1/\btheta^p$, where $p$ is a positive
power, is monotonically increasing in $r$.  However, in an $AdS_2
\times S^2$ spacetime, $\btheta \rightarrow 0$ and thus such a function
will blow up and not equal the entropy of the corresponding extremal
black hole.

It seems puzzling at first that a c-function could arise from the
analysis of second order equations of motion.  As mentioned in the
introduction, the answer to this puzzle lies in the fact that we were
considering solutions which satisfy asymptotically flat boundary
conditions.  Without imposing any boundary conditions, we cannot prove
monotonicity of $c$. But one can use the arguments above to show that
there is at most one critical point of $c$ as long as the region of
spacetime under consideration has no spacetime singularities and also
the scalar fields take non-singular values in moduli space. If the
critical point occurs at $r=r_*$, $c$ monotonically decreases for all
$r<r_*$ and cannot have another critical point. Similarly, for
$r>r_*$. From the Raychaudhuri equation it follows that the critical point, at 
$r_*$, is a maximum. 
 
Usually the discussion of supersymmetric attractors involves the
regions from the horizon to asymptotic infinity.  But we can also ask
what happens if we go inside the horizon. This is particularly
interesting in the non-extremal case where the inside is a time
dependent cosmology. In the supersymmetric case one finds that the
central charge (and its square) has a minimum at the horizon and
increases as one goes away from it towards the outside and also
towards the inside.  This can be seen as follows. Using continuity at
the horizon a modulus take the form in an attractor solution, \be
\label{valphi}
\phi(r)-\phi_0\sim |r-r_H|^\alpha \ee where $\alpha$ is a positive
coefficient and $\phi_0$ is the attractor value for the modulus
\footnote{We are working in the coordinates, eq.(\ref{metric2}). These
  breakdown at the horizon but are valid for $r>r_H$ and also $r<r_H$
  (where $a^2<0$).  The solution written here is valid in both these
  regions; for $r=r_H$ we need to take the limiting value.}.  Since
the central charge is minimised by $\phi_0$, one finds by expanding in
the vicinity of $r=r_H$, that the central charge is also minimised as
a function of $r$ \footnote{The effective potential $V_{eff}$ in the
  non-supersymmetric case is similar.  As a function of $r$ it attains
  a local minimum at the horizon.}.  In contrast, the c-function we
have considered here, monotonically decreases inside the horizon till
we reach the singularity. In fact it follows from the Raychaudhuri
equation that the expansion parameter $\btheta$ monotonically decreases and
becomes $-\infty$ at the singularity.

\section{The c-function In Higher Dimensions}

We analyse higher dimensional generalisations in this section.
Consider a system consisting of gravity, gauge fields with rank $q$
field strengths, $F^a_{m_1 \cdots m_q}, a=1, \cdots N$, and moduli
$\phi^i, i=1, \cdots n$, in $p+q+1$ dimensions, with action,
\begin{equation}
\label{acthd}
S=\frac{1}{\kappa^{2}}\int d^{D}x\sqrt{-G}\left(R-2g_{ij}(\partial\phi^{i})\partial \phi^j 
  -f_{ab}(\phi^{i}){1\over q!} F_{\mu\nu....}^{a}F^{b\ \mu\nu......}\right).
\end{equation}
 
Take a metric and field strengths  of form, 
\begin{eqnarray}
  ds^{2} &  = &  a(r)^{2}\left(-dt^{2} +  \sum_{i=1}^{p-1}dy_{i}^{2}\right) + a(r)^{-2}dr^2 +b(r)^{2}d\Omega_{q}^{2},
  \label{metrichd} \\
  F^{a} & = & Q_{m}^{a}\omega_q \label{fstrengthhd}.
\end{eqnarray}
Here $d\Omega_q^2$ and $\omega_q$ are the volume element and volume
form of a unit $q$ dimensional sphere sphere. Note that the metric has
Poincare invariance in $p$ direction, $t, y_i,$ and has $SO(q)$
rotational symmetry.  The field strengths thread the $q$ sphere and
the configuration carries magnetic charge.  Other generalisations,
which we do not discuss here include, forms of different rank, and
also field strengths carrying both electric and magnetic charge.

Define an effective potential, 
\be
\label{veffhd}
V_{eff}=f_{ab}(\phi^i)Q^a_m Q^b_m.
\ee
 
Now, as we discuss further in appendix \ref{sec:higher-dimensional-p},
it is easy to see that if $V_{eff}$ has a critical point where
$\partial_{\phi^i} V_{eff}$ vanishes, then by setting the scalars to
be at their critical values, $\phi^i=\phi^i_0$, one has extremal and
non extremal black brane solutions in this system with metric,
eq.(\ref{orgsol}).  For extremal solutions, the near horizon limit is
$AdS_{p+1} \times S^q$, with metric given by eq.(\ref{adshd}), 
\be
\label{nh}
ds^2={r^2 \over R^2} \left(-dt^2+dy_i^2\right) + {R^2\over r^2}dr^2 +
b_H^2 d\Omega_q^2 
\ee 
where
\begin{eqnarray} 
\label{valradius}
R & = & \left(\frac{p}{q-1}\right)b_H   \\
(b_H)^{2(q-1)} & = & \frac{p}{(p+q-1)(q-1)} V_{eff}(\phi^i_0).
\end{eqnarray}
In the extremal case, using arguments analogous to \cite{gijt} one can
show that the $AdS_{p+1} \times S^q$ solution is an attractor if the
effective potential is minimised at the critical point $\phi^i_0$.
That is, for  small deviations from the attractor values for the
moduli at infinity, there is an extremal solution in which the moduli
are drawn to their critical values at the horizon and the geometry in
the near-horizon region is $AdS_{p+1} \times S^q$.

We now turn to discussing the c-function in this system.  The
discussion is motivated by the analysis in \cite{fgpw} of a c-theorem
in AdS space.  Our claim is that a c-function for the system under
consideration is given by,
\be
\label{chd}
c  =  c_0 { 1\over {\tilde A}^{(p-1)}}.
\ee 
Here, $c_0$ is a constant of proportionality chosen so that $c>0$.  ${{\tilde A}}$ is defined by
\be
\label{eqadot}
 {{\tilde A}}= A' \left({a\over b^{q \over p-1}}\right) 
\ee
where $A$ is defined to be, 
\be
\label{defA}
A=\ln(ab^{q\over p-1}), 
\ee
and prime denotes derivative with respect to $r$. 
We show below that for any static, asymptotically flat solution of the
form, eq.(\ref{metrichd}), $c$, eq.(\ref{chd}), is a monotonic
function of the radial coordinate.

The key is once again to use the null energy condition.  Consider
the $R_{tt}G^{tt} -R_{rr}G^{rr}$ component of the Einstein equation.
For the metric, eq.(\ref{metrichd}), we get, 
\be
\label{compeq}
-R_{tt}G^{tt} +R_{rr}G^{rr}= a^2 \left[-(p-1) {a^{''} \over a}-q
  {b^{''}\over b}\right]=T_{\mu \nu} \zeta^\mu \zeta^\nu, 
\ee 
where
$(\zeta^t, \zeta^r)$ are the components of a null vector which satisfy
the relation, $(\zeta^t)^2=-G^{tt}, (\zeta^r)^2=G^{rr}$. The null
energy condition tells us that the r.h.s cannot be negative.  For the
system under consideration the r.h.s can be calculated giving, 
\be
\label{condhd}
-(p-1) {a^{''} \over a}-q {b^{''}\over b}=2 g_{ij} \partial_r \phi^i
\partial_r\phi^j. 
\ee 
It is indeed positive, as would be expected since
the matter fields we include satisfy the null energy condition.

From eq.(\ref{condhd}) we find that 
\be
\label{dervad} {d{{\tilde A}} \over dr}= -{a \over b^{q\over
    p-1}}\left[{2\over p-1} g_{ij}\phi^i\phi^j + \right({q\over p-1}+
{q^2\over (p-1)^2}\left)\left({b'\over b}\right)^2\right] ,
\ee 
and thus, ${d{{\tilde A}} \over dr} \leqslant0$.

Now we turn to the monotonicity of $c$.  Consider a solution which
becomes asymptotically flat as $r\rightarrow \infty$.  Then, $a
\rightarrow 1, b \rightarrow r$, as $r\rightarrow \infty$.  It follows
then that ${{\tilde A}} \rightarrow 0^+$ asymptotically. Since, we learn
from eq.(\ref{dervad}) that ${{\tilde A}}$ is a non-increasing function of
$r$ it then follows that for all $r<\infty$, ${{\tilde A}}>0$.  Since,
$a,b>0,$ we then also learn from, eq.(\ref{eqadot}), that $A'>0$ for
all finite $r$.

Next choose a coordinate $y=-r$ which increases as we go in from
asymptotic infinity.  We have just learned that $dA/dy=-A'<0,$ for
finite $r$.  It is now easy to see that 
\be
\label{monca} {dc \over dy}=-(p-1) {a\over b^{q \over p-1}} c 
{dA  \over dy} {1 \over {{\tilde A}^2}} {d{{\tilde A}} \over dr}. 
\ee 
Then given that $a,b>0$, $c>0$, and $dA/dy<0$, ${d{{\tilde A}} \over dr}
\leqslant0$, 
it follows that $dc/dy \leqslant0$, so that the c-function is a
non-increasing function along the direction of increasing $y$. This
completes our proof of the c-theorem.

For a black brane solution the static region of spacetime ends at a horizon, where $a^2$
vanishes. The c-function monotonically decreases from infinity and in the static region
obtains its minimum value at the horizon.  For the extremal black brane the near horizon
geometry is $AdS_{p+1} \times S^q$. We now verify that for $p$ even the $c$ function evaluated
in the $AdS_{p+1} \times S^q$ geometry agrees with the conformal anomaly in the boundary
Conformal Field Theory.  From eq.(\ref{nh}) we see that in $AdS_{p+1} \times S^q$,
\begin{eqnarray}
a' & = & 1/R \\
b & = & {q-1 \over p}  R.
\label{relhhd}
\end{eqnarray}
where $R$ is the radius of the $AdS_{p+1}$. 
Then 
\be
\label{chor}
c \propto {R^{p+q-1} \over G_N^{p+q+1}} \propto {R^{p-1} \over G_{N}^{p+1}} 
\ee 
where
$G_N^{p+q+1}, G_N^{p+1}$ refer to Newton's constant in the $p+q+1$ dimensional spacetime
and the $p+1$ dimensional spacetime obtained after KK reduction on the $S^q$ respectively.
The right hand side in eq.(\ref{chor}) is indeed proportional to the value of the
conformal anomaly in the boundary theory when $p$ is even \cite{Henningson:1998gx}.  By
choosing $c_0$, eq.(\ref{chd}), appropriately, they can be made equal.  Let us also
comment that $c$ in the near horizon region can be expressed in terms of the minimum value
of the effective potential. One finds that
$c \propto (V_{eff}(\phi^i_0))^{(p+q-1) \over 2(q-1)}$, where the critical values for the
moduli are $\phi^i=\phi^i_0$.

A few comments are worth making at this stage.  We have only considered asymptotically
flat spacetimes here. But our proof of the c-theorem holds for other cases as well. Of
particular interest are asymptotically $AdS_{p+1}\times S^q$ spacetime.  The metric in
this case takes the form, eq.(\ref{nh}), as $r\rightarrow \infty$. The proof is very
similar to the asymptotically flat case. Once again one can argue that $A'>0$ for
$r <\infty$ and then defining a coordinate $y=-r$ it follows that $dc/dy$ is a
non-increasing function of $y$.  The c-theorem allows for flows which terminate in another
asymptotic $AdS_{p+1} \times S^q$ spacetime. The second $AdS_{p+1} \times S^q$ space-time,
which lies at larger $y$, must have smaller $c$. Such flows can arise if $V_{eff}$ has
more than one critical point.  It is also worth commenting that requiring that $c$ is a
constant in some region of spacetime leads to the unique solution (subject to the
conditions of a metric which satisfies the ansatz, eq.(\ref{metrichd})) of
$AdS_{p+1} \times S^q$ with the scalars being constant and equal to a critical value of
$V_{eff}$.

We mentioned above that our definition of the $c$ function is motivated by \cite{fgpw}.
Let us make the connection clearer. The c-function in\footnote{Another  c-function has been defined in
 \cite{Nojiri:2000kh}. } \cite{fgpw},\cite{Girardello:1998pd} is defined for a
spacetime of the form,
\be
\label{fc}
ds^2=e^{2A}\sum_{\mu,\nu=0, \cdots p} \eta_{\mu\nu} dy^\mu dy^\nu + dz^2,
\ee
and is given  by  
\be
\label{deffc}
c={c_0 \over (dA/dz)^{p-1}}.
\ee
Note that eq.(\ref{fc}) is the Einstein frame metric in $p+1$ dimensions.  Starting with
the metric, eq.(\ref{metrichd}), and Kaluza-Klein reducing over the $Q$ sphere shows that
$A$ defined in eq.(\ref{defA}) agrees with the definition eq.(\ref{fc}) above and $dA/dz$
agrees with ${{\tilde A}}$ in eq.(\ref{eqadot}). This shows that the c-function
eq.(\ref{chd}) and eq.(\ref{deffc})are the same.

The monotonicity of $c$ follows from that of ${{\tilde A}}$, eq.(\ref{eqadot}).  One can
show that for a congruence of null geodesics moving in the radial direction, with constant
$(\theta,\phi)$, the expansion parameter $\btheta$ is given by,
\be
\label{thd}
\btheta=\left({a'\over a} + {q\over p-1} {b'\over b}\right).
\ee
Raychaudhuri's equation and the null energy condition then tells us that
${d\btheta\over dr} <0$.  However, in an $AdS_{p+1} \times S^q$ spacetime $\btheta$
diverges, this behaviour is not appropriate for a c-function.  From eq.(\ref{eqadot}) we
see that ${{\tilde A}}$ differs from $\btheta$ by an additional multiplicative factor,
$a / b^{q\over p-1}$. This factor is chosen to preserve monotonicity and now ensures that
$c$ goes to a finite constant in $AdS_{p+1} \times S^q$ spacetimes. A similar comment also
applies to the c-function discussed in \cite{fgpw}.

\section{Concluding Comments}

In two-dimensional field theories it has been suggested sometime ago
\cite{Banks:1987qs,Das:1988vd,Vafa:1988ue} that the $c$ function plays the role of a
potential, so that the RG equations take the form of a gradient flow,
$$\beta_i=-{\partial c \over \partial g^i}, $$
where $c$ is the Zamolodchikov c-function~\cite{Zamolodchikov:1986gt}.  This phenomenon
has a close analogy in the case of supersymmetric black holes, where the radial evolution
of the moduli is determined by the gradient of the central charge in a first order
equation.  In contrast, the c-function we propose does not satisfy this property in either
the supersymmetric or the non-supersymmetric case.  In particular, in the
non-supersymmetric case the scalar fields satisfy a second order equation and in
particular the gradient of the c-function does not directly determine their radial
evolution.

It might seem confusing at first that our derivation of the c-theorem followed from the
second order equations of motion.  The following simple mechanically model is useful in
understanding this.  Consider a particle moving under the force of gravity. The c-function
in this case is the height $x$ which satisfies the condition 
\be
\label{he} 
{\ddot x}=-g, 
\ee 
where $g$ is the acceleration due to gravity.  Now, if the initial conditions are such
that ${\dot x}<0$ then going forwards in time $x$ will monotonically decrease.  However,
if the direction of time is chosen so that ${\dot x}>0,$ going forward in time there will
be a critical point for $x$ and thus $x$ will not be a monotonic function of time.  In
this case though there can be at most one such critical point.

While the equations of motion that govern radial evolution are second order, the attractor
boundary conditions restrict the allowed initial conditions and in effect make the
equations first order.  This suggests a close analogy between radial evolution and RG
flow.  The existence of a c-function which we have discussed in this paper adds additional
weight to the analogy.  In the near-horizon region, where the geometry is
$AdS_{p+1} \times S^q$, the relation between radial evolution and RG flow is quite precise
and well known. The attractor behaviour in the near horizon region can be viewed from the
dual CFT perspective.  It corresponds to turning on operators which are irrelevant in the
infra-red. These operators are dual to the moduli fields in the bulk, and their being
irrelevant in the IR follows from the fact that the mass matrix, eq.(\ref{massmatrix}),
has only positive eigenvalues.

It is also worth commenting that the attractor phenomenon in the context of black holes is
quite different from the usual attractor phenomenon in dynamical systems. In the latter
case the attractor phenomenon refers to the fact that there is a universal solution that
governs the long time behaviour of the system, regardless of initial conditions. In the
black hole context a generic choice of initial conditions at asymptotic infinity does not
lead to the attractor phenomenon. Rather there is one well behaved mode near the horizon
and choosing an appropriate combination of the two solutions to the second order equations
at infinity allows us to match on to this well behaved solution at the horizon. Choosing
generic initial conditions at infinity would also lead to triggering the second mode near
the horizon which is ill behaved and typically would lead to a singularity.

Finally, we end with some comments about attractors in cosmology.  Scalar fields exhibit a
late time attractor behaviour in FRW cosmologies with growing scale factor (positive
Hubble constant $H$).  Hubble expansion leads to a friction term in the scalar field
equations, 
\be
\label{ft}  {\ddot\phi}+ 3 H {\dot \phi} + \partial_\phi V=0.  
\ee 
As a result at late times the scalar fields tend to settle down at the minimum of the
potential generically without any precise tuning of initial conditions.  This is quite
different from the attractor behaviour for black holes and more akin to the attractor in
dynamical systems mentioned above.

Actually in $AdS$ space there is an analogy to the cosmological attractor.  Take a scalar
field which has a negative $(mass)^2$ in AdS space (above the BF bound). This field is
dual to a relevant operator.  Going to the boundary of AdS space a perturbation in such a
field will generically die away. This is the analogue of the late time behaviour in
cosmology mentioned above.  Similarly there is an analogue to the black hole attractor in
cosmology. Consider dS space in Poincare coordinates, 
\be
\label{dp}
ds^2=-{dt^2\over t^2} + t^2  dx_i^2,
\ee
and a scalar field with potential $V$ propagating in this background.  Notice that
$t\rightarrow 0$ is a double horizon.  For the scalar field to be well behaved at the
horizon, as $t\rightarrow 0$, it must go to a critical point of $V$, and moreover this
critical point will be stable in the sense that small perturbations of the scalar about
the critical point will bring it back, if $V''<0$ at the critical point, i.e., if the
critical point is a maximum.  This is the analogue of requiring that $V_{eff}$ is at a
minimum for attractor behaviour in black hole
\footnote{The sign reversal is due to the interchange of a space and time directions.}.
It is amusing to note that a cosmology in which scalars are at the maximum of their
potential, early on in the history of the universe, could have other virtues as well in
the context of inflation.

\centerline{\bf Acknowledgements}
\noindent
We thank N. Iizuka for discussion. 
This research is supported by the Government of India. 
S.T. acknowledges support from the Swarnajayanti Fellowship, DST, Govt. of India. 
Most of all we thank the people of India for generously supporting research in String Theory.  
    
\appendix
\renewcommand{\theequation}{A.\arabic{equation}}
\setcounter{equation}{0}
\section{$V_{eff}$ Need Not Be Monotonic}\label{sec:v_eff-need-not}

In this appendix we construct an explicit example showing that $V_{eff}$ as a function of
the radial coordinate need not be monotonic. The basic point in our example is simple.
The scalar field $\phi$ is a monotonic function of the radial coordinate, $r$,
eq.(\ref{metric2}) . But the effective potential is not a monotonic function of $\phi$,
and as a result is not monotonic in $r$.

\begin{figure}[htb]
  \begin{center}
    \includegraphics[%
  scale=0.6]{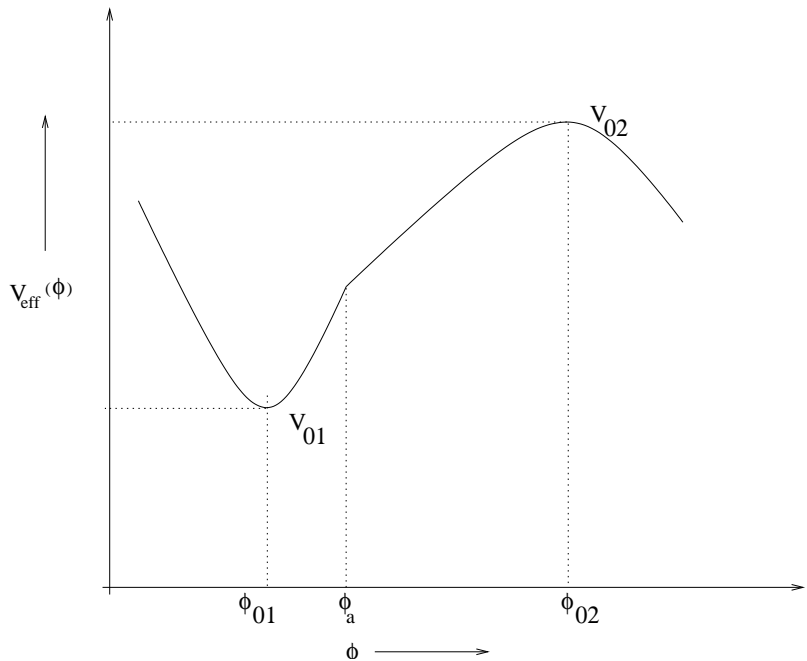}
\end{center}
\caption{The effective potential $V_{eff}$ as a function of $\phi$}
\end{figure}
We work with the following simple $V_{eff}$ to construct such a solution,
\begin{eqnarray}
  \label{veff}
  V_{eff} & = & V_{01} + {1\over 2} m^2 (\phi-\phi_{01})^2, ~\phi\leqslant \phi_a \\
  V_{eff} & = & V_{02} - {1\over 2} m^2 (\phi-\phi_{02})^2, ~\phi\geqslant \phi_a. 
\end{eqnarray}
At $\phi_a$, the potential is continuous, giving the relation, 
\be
\label{phia}
V_{02}=V_{01}+{1\over 2} m^2 (\phi-\phi_{01})^2 + {1\over 2} m^2
(\phi-\phi_{02})^2.  
\ee 
We will take the potential as being specified by $V_{01}, \phi_{01},
\phi_{02}, \phi_{a}, m^2$ with $V_{02}$ being determined by
eq.(\ref{phia}). The effective potential is given in fig. 1. Note that
with a minimum at $\phi_{01}$ and a maximum at $\phi_{02},$ $V_{eff},$
is a non-monotonic function of $\phi$.  Note also that the the first
derivative of the potential has a finite jump at $\phi=\phi_a$.  Since
the equations of motion are second order this means the scalar fields
and the metric components, $a,b$, and their first derivatives will be
continuous across $\phi_a$. The finite jump is thus mild enough for
our purposes.

The attractor value for the scalar is $\phi_{01}$. By setting
$\phi=\phi_{01}$, independent of $r$, we get an extremal Reissner
Nordstrom black hole solution.  The radius of the horizon, $r_H$ in
this solution is given by
\be
\label{valr}
r_H^2=V_{01}.
\ee
This solution is our starting point. We now construct the solution of
interest in perturbation theory,  following the analysis in
\cite{gijt}, whose conventions we also adopt.  For the validity of
perturbation theory, we take, $\phi_a-\phi_{01} \ll 1$, and also
$\phi_{02}-\phi_{01} \ll 1$.  The non-monotonicity of the potential
 then comes into play even when the scalar field makes only  small
excursions around the minimum $\phi_{01}$. In addition we will also
take, ${4m^2\over r_H^2} < 1$, it then follows that ${V_{02}-V_{01}
  \over V_{01}} \ll 1$.
  
We construct the solution for the scalar field to first order in
perturbation theory below. In the solution  the scalar field is a
monotonic function
of $r$.    This allows the solution to be described in two regions.
In
region I, $\phi_{01} \leqslant\phi \leqslant\phi_a$, it is given by,
\begin{eqnarray}
\label{riphi}
\phi & = & \phi_{01}+A(r-r_H)^\alpha, \\
\alpha & = & {1\over 2} \left(\sqrt{1+{4 m^2\over r_H^2}}-1\right).
\end{eqnarray}
And in region II, $\phi>\phi_a$, it is given by,  
\begin{eqnarray}
\label{rtphi}
\phi & = & \phi_{02}+ B_1(r-r_H)^{(-\gamma_1)} + B_2 (r-r_H)^{(-\gamma_2)}, \\
\gamma_1 & = & {1\over 2}\left(1+ \sqrt{1-{4 m^2 \over r_H^2}}\right), \\
\gamma_2 & = & {1\over 2}\left(1- \sqrt{1-{4 m^2 \over r_H^2}}\right).
\end{eqnarray}
The boundary between the two is at $r_a$, where $\phi=\phi_a$, and
$\phi$ and its first derivative with respect to $r$ are continuous.
The continuity conditions allow us to solve for $B_1,B_2,$ in terms of
$A$, and also determine $r_a$ in terms of $A$.  The solution is thus
completely specified by the constant, $A$. $r_a$ satisfies the
relation, 
\be
\label{valra}
\left(1-{r_h\over r_a}\right)^\alpha={(\phi_a-\phi_{01}) \over A}.
\ee

We will omit some details of the subsequent analysis. One finds that
as long as
\be
\label{condA}
(\phi_{a}-\phi_{01})<A < \left({\gamma_1\over \gamma_2}\right)^{\alpha\over \gamma_1-\gamma_2}(\phi_a-\phi_{01}),
\ee
the scalar field monotonically evolves with $r$ and transits from region I to region II as
$r$ increases.  Now we see from eq.(\ref{rtphi}) that if $B_1+B_2>0$,
$\phi(r \rightarrow \infty)>\phi_{02}$.  This ensures that $V_{eff}$ is not a monotonic
function of $r$. It will first increases and then decreases as $r$ decreases from $\infty$
to $r_H$. The condition, $B_1+B_2>0,$ gives rise to the condition,
\be
\label{condc}
(\phi_{02}-\phi_a)<\alpha {[1-(1-{r_H\over r_a})^{\gamma_1-\gamma_2}] \over 
[\gamma_1-\gamma_2(1-{r_H\over r_a})^{\gamma_1-\gamma_2}]}  (\phi_a -\phi_{01}).
\ee
Having picked a value of $A$ that lies in the range, eq.(\ref{condA}),
we can then determine $r_a$ from eq.(\ref{valra}). As long as
$\phi_{02}$ is small enough and satisfies condition eq.(\ref{condc})
we see that the asymptotic value of $\phi(r\rightarrow
\infty)>\phi_{02}$. It then follows, as argued above, that in the
resulting solution $V_{eff}$ is not a monotonic function of $r$.

We end with three comments. First, we have not obtained the the corrections to the metric
components $a,b$ in perturbation theory here. But this can be done following the analysis
in \cite{gijt}. One finds that the corrections are small.  Second, the c-function is of
course monotonic as a function of the radial coordinate in this example too.  The area of
the extremal Reissner Nordstrom black hole monotonically decreases and this is true even
after including the small corrections in perturbation theory.  Finally, we have not
obtained the effective potential above starting with gauge fields coupled to moduli.  In
fact, for dilaton-like couplings, the simplest example we have been able to construct,
where $V_{eff}$ has multicritical points with some minimal and maxima, involves two
moduli, a dilaton and axion, and two gauge fields.  Our discussion above has a close
parallel in this case as well (with both dilaton and axion excited) and we expect, by
dialling the charges and couplings, that the analogue of condition eq.(\ref{condc}) can be
met leading to solutions where $V_{eff}$ evolves non-monotonically with the radial
coordinate $r$.

\section{More Details in Higher Dimensional Case}\label{sec:more-details-higher}
\renewcommand{\theequation}{B.\arabic{equation}}
\setcounter{equation}{0}

The equations of motion that follow from the action, eq.(\ref{acthd}), are,
\be
\begin{array}{ccl}
  R_{\mu\nu}-2\partial_{\mu}\phi_{i}\partial_{\nu}\phi_{i} & = & \frac{q}{q!}f_{ab}(\phi_{i})
  F_{\mu\lambda....}^{a}F_{\nu}^{b\lambda.....}-\frac{q-1}{(p+q-1)q!}G_{\mu\nu}f_{ab}
  (\phi_{i})F_{\mu\nu....}^{a}F^{b\ \mu\nu......}\\
  \frac{1}{\sqrt{-G}}\partial_{\mu}(\sqrt{-G}\partial^{\mu}\phi_{i}) & = & \frac{1}{4q!}\partial_{i}f_{ab}(\phi_{i})
  F_{\mu\nu....}^{a}F^{b\ \mu\nu......}\\
  \partial_{\mu}(\sqrt{-G}f_{ab}(\phi_{i})F^{b\mu\nu}) & = & 0.
\end{array}
\ee

Substituting for the gauge fields from eq.(\ref{fstrengthhd}) we learn
that $R_{tt}={a^2\over b^2} ({q-1\over p}) R_{\theta \theta}$, which
yields the equation,
\begin{equation}
  pb^{2}\left(pa^{'^{2}}+\frac{qaa^{'}b^{'}}{b}+aa^{''}\right)=(q-1)\left((q-1)-(p+1)aba^{'}b^{'}-a^{2}\left((q-1)b^{'2}+bb^{''}\right)\right)
  \label{eq1}
\end{equation}
where we have computed the curvature components using the metric, eq.(\ref{metrichd}).
The $R_{rr}-{\frac{G^{tt}}{G^{rr}}}R_{tt}$ component of the Einstein
equation gives
\begin{equation}
(p-1)\frac{a^{''}}{a}+\frac{qb^{''}}{b}=-2 g_{ij}\partial_r\phi^{i} \partial_r\phi^j. \label{aeq1}
\end{equation}

Also the $R_{rr}$
component itself yields a first order ``energy'' constraint,
\begin{equation}
(p(p-1)b^{2}a^{'2}+2pqaba^{'}b^{'}+q(q-1)(-1+a^{2}b^{'2}))
=2a^{2}b^{2}g_{ij}\partial_{r}\phi^{i}\partial_{r}\phi^{j}-V_{eff}(\phi_{i})b^{-2(q-1)}
\label{rconstraint}
\end{equation}
where $V_{eff}$ is defined in eq.(\ref{veffhd}).

The equation of motion of the scalar field is given by,

\begin{equation}
\partial_{r}(a^{p+1}b^{q}\partial_{r}g_{ij}\phi^{j})=\frac{a^{p-1}\partial_{i}V_{eff}}{4b^{q}}.
\end{equation}

Setting $\phi^i =\phi^i_0$, where $\phi^i_0$ is a critical point of
$V_{eff}$ one finds that $AdS_{p+1} \times S^q$ is a solution of these
equations with metric, eq.(\ref{nh}).

\section{Higher Dimensional $p$-Brane Solutions}\label{sec:higher-dimensional-p}
\renewcommand{\theequation}{C.\arabic{equation}}
\setcounter{equation}{0}

Fixing the scalars at their attractor values, as described in section 4,  we are left with the
action
\begin{equation}
S=\frac{1}{\kappa^{2}}\int d^{D}x\sqrt{-G}\left\{ R-\frac{1}{q!}\sum_{a}{F_{(q)}^{a}}^{2}\right\} 
\label{eq:higher_d_action}
\end{equation}
where $f_{ab}$ has been diagonalised and the attractor values of the
scalars have been absorbed into the a redefinition of the gauge
charges, $Q^{a}$. We denote the new charges as $\bar{Q}^{a}$.

To find solutions, we can dimensionally reduce this action along the
brane and use known blackhole solutions. To this end take the metric
\begin{equation}
  ds^{2}=\underbrace{e^{\lambda\rho}d\hat{s}^{2}}_{t,r,\omega_{1},\ldots,\omega_{q}}
  +\underbrace{e^{-\left(\frac{q}{p-1}\right)\lambda\rho}dy^{2}}_{i_{1}\ldots i_{p-1}}
\label{eq:dim:red:metric}
\end{equation}
where
\begin{equation}
\lambda=\pm\sqrt{\frac{2(p-1)}{q(p+q-1)}}
\end{equation}
 then 
\begin{eqnarray}
R & = & e^{-\lambda\rho}\left(\hat{R}-\lambda^{2}\hat{\nabla}^{2}\rho-\frac{1}{2}(\hat{\nabla}\rho)^{2}\right)
\end{eqnarray}
where $\hat{R}$ and $\hat{\nabla}$ are respectively the Ricci scalar
and covariant derivative for $d\hat{s}^{2}$. The coefficient,
$\lambda$, has been fixed by requiring that, we remain in the Einstein
frame, and that the kinetic term for $\rho$ has canonical
normalisation.  Upon neglecting the boundary term, the action becomes
\begin{equation}
  S=\frac{V_{(p-1)}}{\kappa^{2}}\int d^{(q+2)}x\sqrt{-\hat{G}}
  \left\{ 
    \hat{R}-\frac{1}{2}\left(\hat{\nabla}\rho\right)^{2}
    -\frac{1}{q!}e^{\beta\rho}\sum_{a}{\left({\hat{F}}_{(q)}^{a}\right)}^{2}
  \right\} 
\label{eq:lower_d_action}
\end{equation}
where 
\begin{equation}
\beta=-(q-1)\lambda.
\end{equation}
The black hole solution to eq.(\ref{eq:lower_d_action}) is \cite{Gibbons:1987ps,Horowitz:1991cd}:
\begin{eqnarray}
  d\hat{s}^{2} & = & 
  -\left(f_{+}\right)\left(f_{-}\right)^{1-\hat{\gamma}(q-1)}dt^{2}
  +\left(f_{+}\right)\left(f_{-}\right)^{\hat{\gamma}-1}du^{2}+\left(f_{-}\right)^{\hat{\gamma}}u^{2}d\Omega_{q}^{2}\\
  e^{\lambda\rho} & = & \left(f_{-}\right)^{-\hat{\gamma}}\\
  f_{\pm} & = & \left(1-\left(\frac{u_{\pm}}{u}\right)^{q-1}\right)
\end{eqnarray}
where 
\begin{equation}
  \hat{\gamma}=\frac{2(p-1)}{(q-1)p}
\end{equation}
with
\begin{eqnarray}
  {\hat F}^a &=& {\bar Q}^a \omega_q \\
  \sum_{a}{(\bar{Q}^{a})}^2&=&\frac{\hat{\gamma}(q-1)^{3}(u_{+}u_{-})^{q-1}}{\beta^{2}}.
\label{eq:veff:soln}
\end{eqnarray}
Using eq.(\ref{eq:dim:red:metric}) we find the solution to the original
action, eq.(\ref{eq:higher_d_action}), is 
\begin{eqnarray}
\label{orgsol}
ds^{2} & = & 
\left(f_{-}\right)^{\frac{2}{p}}\left(-\left(\frac{f_{+}}{f_{-}}\right)dt^{2}+dy^{2}\right)
+\left(f_{+}f_{-}\right)^{-1}du^{2}+u^{2}d\Omega_{q}^{2}.
\end{eqnarray}
So finally, the extremal solution is 
\begin{eqnarray}
\label{eq:extremal:sol} 
ds^{2} & = & \left(f\right)^{\frac{2}{p}}\left(-dt^{2}+dy^{2}\right)+\left(f\right)^{-2}du^{2}+u^{2}d\Omega_{q}^{2}\\
f & = & \left(1-\left(\frac{b_{H}}{u}\right)^{q-1}\right)
\end{eqnarray}
where $b_{H}=u_{\pm}$. Now we take the near horizon limit,
\begin{eqnarray}
  u & \stackrel{\epsilon\rightarrow0}{\longrightarrow} & b_{H}+\epsilon R\left(\frac{r}{R}\right)^{p},
\end{eqnarray}
with $t$ and $y$ rescaled appropriately, which indeed gives the
near horizon geometry $AdS_{p+1}\times S^{q}$: 
\begin{eqnarray}
\label{adshd}
ds^{2} & = & \frac{r^2}{R^2}\left(-dt^{2}+dy^{2}\right)+\frac{R^2}{r^2}dr^{2}+b_{H}^{2}d\Omega_{q}^{2}
\end{eqnarray}
where
\begin{equation}
  R=\left(\frac{p}{q-1}\right)b_{H}
\end{equation}
and
\begin{equation}
  V_{eff}\stackrel{\mathrm{eq.}(\ref{eq:veff:soln})}{=}\frac{(p+q-1)(q-1)}{p}(b_{H})^{2(q-1)}.
\end{equation}



\begin{thebibliography}{99}
\addcontentsline{toc}{section}{References}

\bibitem{Ferrara:1995ih} S.~Ferrara, R.~Kallosh and A.~Strominger,
  Phys.\ Rev.\ D {\bf 52}, 5412 (1995) [arXiv:hep-th/9508072].



\bibitem{Cvetic:1995bj}
  M.~Cvetic and A.~A.~Tseytlin,
  Phys.\ Rev.\ D {\bf 53}, 5619 (1996)
  [Erratum-ibid.\ D {\bf 55}, 3907 (1997)] [arXiv:hep-th/9512031].



\bibitem{Strominger:1996kf} A.~Strominger,
  Phys.\ Lett.\ B {\bf 383}, 39 (1996) [arXiv:hep-th/9602111].



\bibitem{Ferrara:1996dd} S.~Ferrara and R.~Kallosh,
  Phys.\ Rev.\ D {\bf 54}, 1514 (1996) [arXiv:hep-th/9602136].



\bibitem{Ferrara:1996um} S.~Ferrara and R.~Kallosh,
  Phys.\ Rev.\ D {\bf 54}, 1525 (1996) [arXiv:hep-th/9603090].




\bibitem{Cvetic:1996zq}
  M.~Cvetic and C.~M.~Hull,
  Nucl.\ Phys.\ B {\bf 480}, 296 (1996) [arXiv:hep-th/9606193].



\bibitem{Ferrara:1997tw} S.~Ferrara, G.~W.~Gibbons and R.~Kallosh,
  Nucl.\ Phys.\ B {\bf 500}, 75 (1997) [arXiv:hep-th/9702103].




\bibitem{Gibbons:1996af} G.~W.~Gibbons, R.~Kallosh and B.~Kol,
  Phys.\ Rev.\ Lett.\ {\bf 77}, 4992 (1996) [arXiv:hep-th/9607108].



\bibitem{Denefa} F.~Denef,
  JHEP {\bf 0008}, 050 (2000) [arXiv:hep-th/0005049].



\bibitem{Denef:2001xn} F.~Denef, B.~R.~Greene and M.~Raugas,
  JHEP {\bf 0105}, 012 (2001) [arXiv:hep-th/0101135].





\bibitem{Ooguri:2004zv} H.~Ooguri, A.~Strominger and C.~Vafa,
  Phys.\ Rev.\ D {\bf 70}, 106007 (2004) [arXiv:hep-th/0405146].



\bibitem{LopesCardoso:1998wt} G.~Lopes Cardoso, B.~de Wit and
  T.~Mohaupt,
  Phys.\ Lett.\ B {\bf 451}, 309 (1999) [arXiv:hep-th/9812082].

\bibitem{LopesCardoso:1999ur}
  G.~Lopes Cardoso, B.~de Wit and T.~Mohaupt,
  Nucl.\ Phys.\ B {\bf 567}, 87 (2000)
  [arXiv:hep-th/9906094].



\bibitem{Dabholkar:2004yr} A.~Dabholkar,
  [arXiv:hep-th/0409148].





\bibitem{Ooguri:2005vr} H.~Ooguri, C.~Vafa and E.~P.~Verlinde,
  [arXiv:hep-th/0502211].



\bibitem{Dijkgraaf:2005bp} R.~Dijkgraaf, R.~Gopakumar, H.~Ooguri and
  C.~Vafa,
  [arXiv:hep-th/0504221].



\bibitem{Sen:2005wa} A.~Sen,
  [arXiv:hep-th/0506177].



\bibitem{gijt}
  K.~Goldstein, N.~Iizuka, R.~P.~Jena and S.~P.~Trivedi,
  [arXiv:hep-th/0507096].


\bibitem{Kallosh:2005bj}
  R.~Kallosh,
  [arXiv:hep-th/0509112].



\bibitem{Kallosh:2005ax}
  R.~Kallosh,
  [arXiv:hep-th/0510024].




\bibitem{TTNSA}
  P.~K.~Tripathy and S.~P.~Trivedi,
  [arXiv:hep-th/0511117].


\bibitem{Giryavets:2005nf}
  A.~Giryavets,
  [arXiv:hep-th/0511215].


\bibitem{Kraus:2005vz} P.~Kraus and F.~Larsen,
  [arXiv:hep-th/0506176].


\bibitem{Kraus:2005zm}
  P.~Kraus and F.~Larsen,
  [arXiv:hep-th/0508218].




\bibitem{fgpw}
  D.~Z.~Freedman, S.~S.~Gubser, K.~Pilch and N.~P.~Warner,
  Adv.\ Theor.\ Math.\ Phys.\  {\bf 3}, 363 (1999)
  [arXiv:hep-th/9904017].

\bibitem{Girardello:1998pd}
  L.~Girardello, M.~Petrini, M.~Porrati and A.~Zaffaroni,
  JHEP {\bf 9812}, 022 (1998)
  [arXiv:hep-th/9810126].

\bibitem{Nojiri:2000kh}
  S.~Nojiri, S.~D.~Odintsov and S.~Ogushi,
  Phys.\ Rev.\ D {\bf 62}, 124002 (2000)
  [arXiv:hep-th/0001122].


\bibitem{Sahakian:1999bd}
  V.~Sahakian,
  Phys.\ Rev.\ D {\bf 62}, 126011 (2000)
  [arXiv:hep-th/9910099].




\bibitem{Alvarez:2000jb}
  E.~Alvarez and C.~Gomez,
  [arXiv:hep-th/0009203].



\bibitem{Henningson:1998gx}
  M.~Henningson and K.~Skenderis,
  JHEP {\bf 9807}, 023 (1998)
  [arXiv:hep-th/9806087].




\bibitem{Banks:1987qs}
  T.~Banks and E.~J.~Martinec,
  Nucl.\ Phys.\ B {\bf 294}, 733 (1987).

\bibitem{Das:1988vd}
  S.~R.~Das, G.~Mandal and S.~R.~Wadia,
  Mod.\ Phys.\ Lett.\ A {\bf 4}, 745 (1989).

\bibitem{Vafa:1988ue}
  C.~Vafa,
  Phys.\ Lett.\ B {\bf 212}, 28 (1988).

\bibitem{Zamolodchikov:1986gt}
  A.~B.~Zamolodchikov,
  JETP Lett.\  {\bf 43}, 730 (1986)
  [Pisma Zh.\ Eksp.\ Teor.\ Fiz.\  {\bf 43}, 565 (1986)].


\bibitem{Gibbons:1987ps}
  G.~W.~Gibbons and K.~i.~Maeda,
  Nucl.\ Phys.\ B {\bf 298}, 741 (1988).

  


\bibitem{Horowitz:1991cd}
Gary~T. Horowitz and Andrew Strominger.
\newblock Black strings and p-branes.
\newblock {\em Nucl. Phys.}, B360:197--209, 1991.

\end{thebibliography}
\end{document}